# Schematic Representation Method for Quantum Circuits: An Intuitive Approach to Quantum Gate Effects

**Serkan Akkoyun[1]**

[1] Department of Physics, Faculty of Sciences, Sivas Cumhuriyet University, Sivas, Türkiye

E-mail: sakkoyun@cumhuriyet.edu.tr

**Abstract**

In quantum circuits, qubits and the quantum gates acting on them have traditionally been analysed using matrix algebra and Dirac notation. While powerful, these can be unintuitive for conceptual understanding and rapid problem solving. In this work, a new schematic representation method is developed that visualizes the effects of quantum gates on qubits without relying on complex mathematical operations. In the new notation, quantum bits (qubits) are represented using black ($|0\rangle$) and white ($|1\rangle$) circles. When a quantum gate is applied to a qubit, the circle representing the qubit is visually modified. For example, a Hadamard gate transforms a solid black or white circle into a half-black and half-white circle representing superposition. The work shows how this method simplifies the visualization of quantum algorithms, entanglement, and multi-qubit operations. Thus, the effects of quantum circuits can be analysed with simple schematic representations without having to go into complex mathematical operations.

## 1. Introduction

Quantum computational theory exploits the principles of quantum physics to perform complex computations more efficiently than classical computers. These principles are quantum superposition and quantum entanglement, which have revolutionized computational science [1]. Quantum circuits, which are constructed by arrays of quantum gates implemented on qubits, are fundamental to quantum computation. These quantum circuits are analyzed using matrix algebra and Dirac notation. Visualizations such as the Bloch sphere, which are widely used in quantum computation, are useful for single-qubit states, but they fail to represent multi-qubit interactions and entanglement [2]. Moreover, this representation cannot explicitly depict the evolving quantum state of the system [3]. Therefore, the use of intuitive and simplifying schematic representations that facilitate analysis in quantum circuits is important for understanding the subject [4].

The quantum computing journey was started by Feynman in the early 1980s [5]. Following this beginning, significant advances occurred in quantum algorithms [6-16]. Quantum computers process information using the laws of quantum mechanics, with a completely different approach than the working principles of classical computers. According to quantum mechanics, instead of particles being in a single and specific state, they can be in more than one state at the same time (superposition), and particles that are far away from each other can instantly interact with each other (entanglement). Quantum computers are powerful machines far beyond the limits of classical computers due to these two important principles of quantum mechanics. While bits, the smallest unit that processes in classical computers, can only take the value 0 or 1, quantum bits (qubits) used in quantum computers can be in both $|0\rangle$ and $|1\rangle$ states at the same time due to the quantum superposition feature [17]. Thanks to the quantum entanglement created between qubits, the states of qubits that are far away from each other can be instantly connected to each other. These two quantum physics features allow quantum computers to solve certain computational problems much faster than classical computers. However, these powerful features offered by quantum computers also bring with them a great deal of abstraction and mathematical complexity that is difficult to understand [18].

Although there are studies in the literature to make quantum computing understandable, studies on its visualization are quite limited. In the study conducted by Satanassi et al [19], the process of designing and implementing teaching materials on quantum computing for secondary school students was discussed. The focus of the study is the educational approach developed to teach the quantum teleportation protocol. In the study, the teleportation protocol is explained by making it suitable for high school students. The study by Angara et al [20] describes a series of workshops designed and implemented to teach quantum computing to high school students. The paper examines innovative methods used to teach the fundamental concepts of quantum computing, particularly "unplugged" activities without using a computer and programming approaches using Qiskit [21] and Jupyter notebooks [22] on the IBM Quantum Experience platform [23]. In the study conducted by Sun et al [24], an accessible and applied quantum computing education model for high school students was presented. It was aimed to provide students with a better understanding of the subject by providing hands-on training with a portable user interface and operating system used in quantum computers. Darienzo and Kelly [25] conducted a literature review on quantum information science and technology (QIST) education programs for high school students. The study analyzed studies examining QIST programs for high school students between 2019 and 2023. In the study conducted by Hughes et al [26], an educational program developed to teach quantum computing to high school students was discussed. The course, prepared by the Fermilab Theoretical Physics Department and other educational institutions, was designed for high school students between the ages of 15-18 and its effectiveness was tested and evaluated. In the article published by Dündar-Coecke et al [27], a new educational approach called Quantum Picturalism was introduced and aimed to teach quantum theory to high school students. The basis of the method lies in teaching quantum theory and calculations using visual diagrams. The article examined how intuitive and visual expressions can be more effective than traditional symbolic mathematical operations. The study by Chiofalo et al. [28] focuses on quantum programming education for high school students, using a game-based educational approach to develop an understanding of fundamental concepts such as quantum properties of objects, superposition, entanglement, and measurement.

In this study, we developed a notation that will schematically represent the effects of quantum gates on quantum bits. We represented quantum bits (qubits) by circles. We defined how the colors and patterns of these circles, which can be completely or partially black or white, change with quantum gates. We analyzed example quantum circuits that are mathematically complex to understand with the schematic representation we developed. The approach we develop provides an intuitive way to track state evolution in quantum circuits without the need for complex matrix calculations and mathematical operations. We thus show how quantum computational concepts can be facilitated and how they can serve as an educational tool for students and researchers. In Section II, we gave the basics of the quantum mechanical principles needed for the quantum computations. In Section III, the developed schmetatic representation and notations have been introduced. In Chapter IV, the solution of Bell state generation, super dense coding and quantum teleportation circuits are analyzed using the schematic representation method. Finally in the Conclusion section, the method has been discussed and summarized.

## 2. Theoretical Background for Quantum Computing

### 2.1 Quantum superposition

According to classical physics, an object is in a certain position and is moving at a certain speed, meaning that the visible world around us operates according to clear and definite rules. A ball thrown upwards is either in the air or has fallen to the ground, a light is either on or off. In the quantum world, on the other hand, things are strange beyond our perception. One of the most interesting of these strangenesses is quantum superposition. Quantum phenomena and their effects gain meaning and manifest themselves in the world of atoms and subatomic particles. Things in this world do not follow the logic of "either this or that" as in the macro world we are accustomed to. According to quantum superposition, a subatomic particle can be in more than one state at the same time. For example, an electron in a box should be either on the right or the left of the box, but according to quantum mechanics, this electron can be on both the right and the left of the box at the same time until it is measured. In other words, it cannot be said that the electron is at a certain point, because it contains many possibilities at the same time.

To illustrate by a different example, let's recall that quantum particles like electrons have a property called spin. Spin has two fundamental states, spin-up and spin-down. In classical physics, an electron is either in a spin-up or spin-down state. However, according to quantum superposition, an electron can be in a mixture of spin-up and spin-down states at the same time. One of the thought experiments that best explains the concept of superposition is the Schrödinger's cat experiment. In this experiment, there is a box containing a cat, a radioactive atom, and a poisonous mechanism. If the atom decays, the mechanism works and the cat dies, whereas if the atom does not decay, the cat lives on. However, according to quantum superposition, until the measurement is made, the atom is in a superposition of both decayed and undecayed states. Therefore, the unobserved cat in the closed box can be both dead and alive at the same time.

Therefore, a quantum system in superposition can be in more than one state at a time for a given property. None of these states are definitively true until a measurement (observation) is made, meaning they all exist as potential. This seemingly contradictory situation is one of the fundamental properties of superposition. In the real world, superposition helps us understand how electrons behave inside atoms, how computers work, and even how quantum computers process information. This situation, which we do not notice in our daily lives, is one of the most interesting and revolutionary features of quantum physics.

*2.2 Quantum entanglement*

The strangeness of the quantum world is not limited to superposition. Another strangeness is the phenomenon of quantum entanglement. Accordingly, no matter how far apart two or more quantum particles are from each other, the effect on one will instantly affect the other. In entanglement, two or more particles can superimpose themselves together and have a common uncertain state instead of separate uncertain states.

Entanglement is a phenomenon that cannot be explained by the laws of classical physics. Because in order for two objects to affect each other, they must somehow communicate, and this communication takes time. However, when two electrons are entangled, for example, when the spin of one is measured, the spin of the other is also determined instantly. This phenomenon surprised even Albert Einstein, who called it "spooky interaction at a distance" because it contradicted the law of special relativity. For this reason, Einstein and his colleagues suggested that quantum mechanics was incomplete and that entanglement could actually be explained by unknown hidden variables. However, in the mid-20th century, physicist John Bell put forward Bell's Inequalities, which showed that entanglement could not have a classical explanation. A series of experiments subsequently confirmed that entanglement existed, contrary to Einstein's prediction. Accordingly, quantum entanglement is not just a theoretical phenomenon, but a reality that forms the basis of quantum computers and quantum communication today. In particular, in quantum encryption, it has become possible to transfer data completely securely using two entangled particles.

## 3. Schematic Representations of Qubits and Gates

Among the most striking advantages provided by quantum mechanics are information transmission protocols such as superdense coding and quantum teleportation. These methods provide more efficient and secure information transmission by using the advantages provided by quantum entanglement compared to classical information communication.

In Super Dense Coding, by increasing the capacity of classical communication channels due to quantum entanglement, two bits of classical information can be transmitted over a single qubit using a previously shared entangled quantum state. However, in classical communication, as is known, two bits of physical resources are required to transmit two bits of information. For example, Alice can send two bits of information to Bob using a previously shared Bell state ($\Phi^+ = (|00\rangle + |11\rangle)/\sqrt{2}$). Alice can change the state of the qubit by applying certain quantum gates (Pauli-X or/and Pauli-Z) to her qubit and then send this qubit to Bob. In this way, Bob obtains the correct message that Alice wants to transmit by performing measurements on the qubits.

Quantum Teleportation is a protocol that enables the transfer of quantum information to a distant point by overcoming the quantum no-cloning theorem, which states that information cannot be directly copied through classical channels. In this method, a quantum state is transferred to another particle without physically transmitting it. The teleportation process is performed by a combination of entangled qubit pairs and classical information transmission. Alice performs a measurement using a qubit that has the quantum state she wants to transmit, using one of the entangled qubits she previously shared with Bob. Then, Alice transmits the information she obtains as a result of the measurement to Bob via a classical channel. Bob applies a suitable quantum gate to his qubit using this information and thus obtains the exact quantum state that Alice wants to transmit to him. With the new schematic representations presented in this study, Bell States, Super Dense Coding and Quantum Teleportation circuits are analyzed.

In the schematic representation developed within the scope of this study, the $|0\rangle$ qubit state is represented by a black filled circle (Fig.1.a), and the 1 qubit state is represented by a white empty circle (Fig.1.b), as shown below.

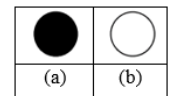

**Fig.1** Qubit states. Black filled circle for $|0\rangle$ (a) and white empty circle for $|1\rangle$ (b)

Although readers might be familiar to the basic information that will be mentioned in this paragraph, it is thought to be useful to remind these once again in order to fully express the newly developed schematic representation. The mathematical representations of the fundamental gates have been given in Table 1.

**Table 1** Effects of some of the fundamental gates on qubit states

| | |
|---|---|
| $H\|0\rangle \to \frac{1}{\sqrt{2}}(\|0\rangle + \|1\rangle)$ | $H\|1\rangle \to \frac{1}{\sqrt{2}}(\|0\rangle - \|1\rangle)$ |
| $X\|0\rangle \to \|1\rangle$ | $X\|1\rangle \to \|0\rangle$ |
| $Y\|0\rangle \to i\|1\rangle$ | $Y\|1\rangle \to -i\|0\rangle$ |
| $Z\|0\rangle \to \|0\rangle$ | $Z\|1\rangle \to -\|1\rangle$ |
| $CNOT\|00\rangle \to \|00\rangle$ | $CNOT\|11\rangle \to \|10\rangle$ |
| $CNOT\|01\rangle \to \|01\rangle$ | $CNOT\|10\rangle \to \|11\rangle$ |

Hadamard gate (H) converts a black circle ($|0\rangle$) or a white circle ($|1\rangle$) into a circle that is half black and half white, visually indicating a superposition state. Here, dividing the circles exactly in half indicates the coefficient of 1/2 of the probability. If any qubit state is preceded by a negative sign, that part of the circle becomes a mirror image on the horizontal axis as in shown below for $H|1\rangle$ operation (Fig.2.a). Pauli-X gate represents a bit-flip operation, swapping black ($|0\rangle$) with white ($|1\rangle$) and vice versa (Fig.2.b). The Pauli-Y gate is an operation that adds a phase factor (*i* coefficient) when flipping the state of a qubit. In the schematic representation, the *i* coefficient is written inside the relevant parts of the circle (Fig.2.c). Pauli-Z gate converts white empty circle ($|1\rangle$) to negative white empty circle ($-|1\rangle$), while leaving black ($|0\rangle$) unchanged (Fig.2.d). The CNOT gate uses interconnected visual elements to indicate conditional state changes. If the control qubit circle part is in the filled black state ($|0\rangle$), it does not change the target qubit circle, while if the control qubit circle part is in the white state ($|1\rangle$), it flips the target qubit part (Fig.2.e).

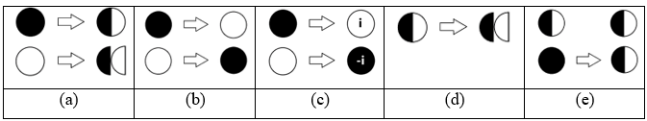

**Fig.2** Effects of the H (a), X (b), Y (c), Z (d) and CNOT (e) gates on the qubit states in the schematic representation

Since a qubit can only take one of two states, $|0\rangle$ and $|1\rangle$, in the schematic representation developed within the scope of this study, the left and right halves of the circles are considered as separate parts. The left and right parts of the each qubit match each other. By combining the left and right semicircles, a full circle is obtained. If a qubit is in the pure $|0\rangle$ ($|1\rangle$) state, both the left and right semicircles will be black (white) and a full black (white) circle will be formed. If a qubit is in the superposition of $|0\rangle$ and $|1\rangle$, its left half will be black and its right half will be white (Fig. 3.a). If there are two qubits in the system and these qubits are entangled with each other, they are shown as in Fig. 3.b. This representation indicates that if the first qubit is in the black state, the second qubit will also be in the black state, and if it is in the white state, the second qubit will also be in the white state. In the schematic representation, the left and right sides of the qubits are paired together. However, if there is no entanglement between these two qubits, the representation will be different as shown in Fig.3.c. Because if the first qubit is in the black (white) state, the second qubit can be in either black or white state. Therefore, while the left half of the first qubit is black, there is no guarantee that the left half of the second qubit will be black or white. The left half of the second qubit can be black or white, while the right half can be white or black for the same reason. The total black and white parts of the circle still remains 1/2.

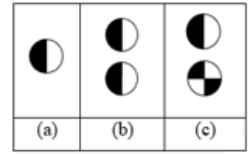

**Fig.3** One qubit in a superposition state (a), two entangled qubits (b) and two seperate qubits in superposition states (c)

## 4. Application of the Schematic Representation

### 4.1 Bell State ($\Phi^+$)

A Bell circuit is a basic quantum circuit used to create quantum entanglement between two qubits. Specifically, the Bell circuit is used to generate Bell states (also known as EPR pairs) as in given in Eqs.1-4. These states form the basis of many quantum computing protocols, such as superdense coding, quantum teleportation, and quantum cryptography. In the first stage of the schematic representation application, the Bell circuit (Figure 4.a) that produces the first of the Bell states ($\Phi^+$) is used to obtain superposition and entanglement between two qubits. The newly developed schematic representation solution of this circuit is shown in Figure 4.b.

$$\Phi^+ = (|00\rangle + |11\rangle)/\sqrt{2} \quad (1)$$
$$\Phi^- = (|00\rangle - |11\rangle)/\sqrt{2} \quad (2)$$
$$\Psi^+ = (|01\rangle + |10\rangle)/\sqrt{2} \quad (3)$$
$$\Psi^- = (|01\rangle - |10\rangle)/\sqrt{2} \quad (4)$$

This state ensures that the measurement result of the two qubits is completely entangled. For example, if one of the qubits in the $|\Phi^+\rangle$ state is measured to find $|0\rangle$, the other qubit will definitely be $|0\rangle$. Similarly, if one is measured to find $|1\rangle$, the other will definitely be $|1\rangle$. As seen in Figure 4, the H gate is first applied to the $q_0$ qubit state and the circle representing this qubit is reformatted so that the left side is black and the right side is white. Then, by the CNOT gate applied between two qubits (q0 and q1), the black left half of $q_0$ did not change the left half of $q_1$, while the white right half of $q_0$ caused the right half of $q_1$ to flip and become white. Thus, two entangled qubits are obtained.

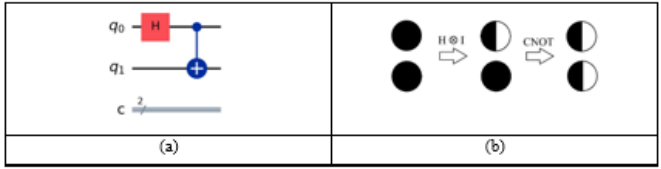

**Fig.4** The quantum circuit (a) for first Bell state and the schematic solution of the circuit (b)

## 4. 2 Super Dense Coding

Super dense coding is a technique where one can send 2 bits of classical information using a quantum system of one qubit as mentioned in Section III. Normally, one can send 1 bit of information over a classical channel. But with super dense coding, it is possible to send 2 bits of information using one qubit. Alice and Bob initially share a pair of qubits that are entangled with each other. Alice encodes the 2 bits of information she wants to send by applying a specific quantum gate to her own qubit. Alice sends the qubit she changed to Bob. Bob measures his own qubit and the incoming qubit together to extract 2 bits of classical information. Whereas in the classical system, you need to send 2 separate bits to send 2 bits. This makes quantum communication more efficient than classical one. It is an important technique especially for quantum internet, secure communication and future quantum networks.

The quantum circuits to be used according to the message Alice wants to send to Bob are shown in Figures 5-8. In each circuit, Alice's qubit ($q_0$) and Bob's qubit ($q_1$) are first entangled with each other in a way that they create a Bell state. For this purpose, Alice first applies the H gate to her own qubit. In the schematic representation, Alice's initially solid black circle becomes half black and half white. In other words, Alice's qubit is now in a superposition state. Then Alice applies the CNOT gate and Bob's black circle also becomes half black and half white. Because, while the full side of Bob's circle corresponding to the black left half of Alice's circle does not change, the half of Bob's circle corresponding to the white right half of Alice flips and becomes white.

### 4.2.1 I gate for sending |00⟩

If the message to be sent is |00⟩, Alice applies the I gate to her qubit. Since this gate will not cause any change in Alice's qubit, the left half of Alice's circle will remain black and the right half will remain white (Fig.5). After applying this gate, Alice sends her qubit to Bob through a quantum channel. Bob applies the CNOT gate to Alice's qubit. The black left half of Alice's circle does not change the left half of Bob's circle, while the white right half causes the right side of Bob's circle to become black. Now the schematic representation of Bob's qubit is a solid black circle. Finally, Bob applies the H gate to Alice's qubit. This circle, which is half black and half white, becomes a solid black circle. If Bob measures both qubits, he will measure two |0⟩ qubit states. Because in the last case, the circles representing Alice's and Bob's qubits are solid black circles.

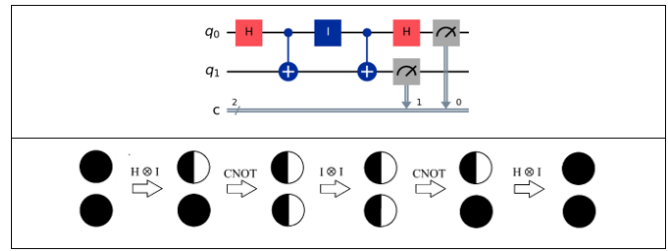

**Fig.5** Quantum circuit sending |00⟩ message by super dense coding (top) and solution of the circuit with schematic representation (bottom)

### 4.2.2 X gate for sending |01⟩

If the message to be sent is |01⟩, Alice applies a Pauli-X gate to the qubit before sending it to Bob. This gate flips Alice's qubit, causing the left half, which is black, to turn white and the left half, which is white, to turn black in the schematic (Fig.6). After Alice applies this gate, she sends her qubit to Bob via any quantum channel. Bob now has both his and Alice's qubits and applies a CNOT to Alice's. The white left half of Alice's circle flips the left half of Bob's circle to black. Since the right half of Alice's qubit is black in the schematic, the right half of Bob's circle remains white. At this stage, the schematic representation of Bob's qubit is a hollow white circle. In the final stage, Bob applies a H gate to Alice's qubit. Alice's qubit state, which is half black and half white, becomes a solid black circle. If Bob were to measure both qubits, the state he would get would be |01⟩, which is the message Alice wanted to send.

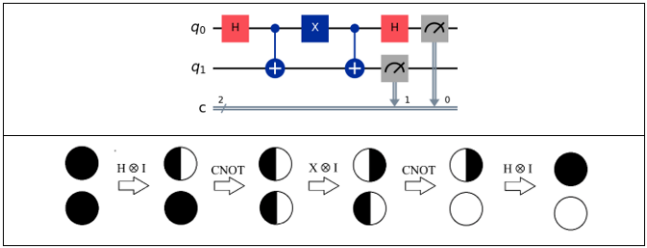

**Fig.6** Quantum circuit sending |01⟩ message by super dense coding (top) and solution of the circuit with schematic representation (bottom)

### 4.2.3 Z gate for sending |10⟩

If the two-bit message Alice wants to send to Bob is |10⟩, Alice applies a Pauli-Z gate to her own qubit and then sends it to Bob. This gate leaves the black left half of Alice's qubit unchanged in the schematic representation, while rotating the white right half in a mirror image on the horizontal axis, as shown in the Fig.7. Alice then transmits her qubit to Bob via a quantum channel. Bob, who has two qubits, applies a CNOT gate to Alice's. Since the black left half of Alice's circle does not cause any change in the left half of Bob's circle, the left side of the schematic circle representing Bob's qubit remains black. However, since the right half of Alice's qubit in the schematic representation is white, the right half of Bob's circle changes from white to black. At this stage, the schematic representation of Bob's qubit is a solid black circle. Bob can now apply the H gate to Alice's qubit. Alice's qubit state, which is half black and half white in a mirror image (-1), becomes a hollow white circle. Ultimately, if Bob measures both qubits, the qubits he will obtain, and therefore the message he will receive, will be in the form of |10⟩.

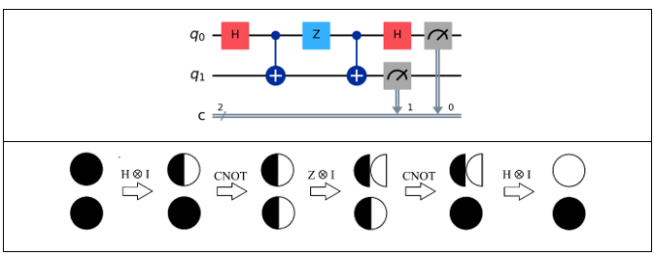

**Fig.7** Quantum circuit sending |10⟩ message by super dense coding (top) and solution of the circuit with schematic representation (bottom)

### 4.2.4 ZX gate for sending |11⟩

In the last case we will examine in super dense coding, suppose Alice wants to send the message |11⟩ to Bob. In this case, Alice first applies a Pauli-X gate to her own qubit, and then immediately a Pauli-Z gate. After applying the Pauli-X gate, the left half of Alice's qubit, which is black, is flipped to white, and the right half, which is white, is flipped to black, as shown in the schematic (Fig.8). By applying the Pauli-Z gate to this last qubit case, Alice causes the left half of her qubit, which is white, to rotate horizontally in a mirror image (-1), while the right half, which is black, remains unchanged. After applying these two gates in succession, Alice transmits her qubit to Bob via the quantum channel. At this stage, Bob applies a CNOT gate to Alice's qubit. Thus, the left half of Alice's circle, which is white, flips the left half of Bob's circle from black to white. Since the right side of Alice's schematic qubit is black, the right side of Bob's schematic qubit remains white without any change. After this step, the schematic representation of Bob's qubit is a hollow white circle. Bob applies the H gate to Alice's qubit and Alice's qubit state, which is half black and half white in a mirror image (-1), becomes a hollow white circle. As a result, Bob measures both qubits and the result he gets is |11⟩, the message Alice wants to send.

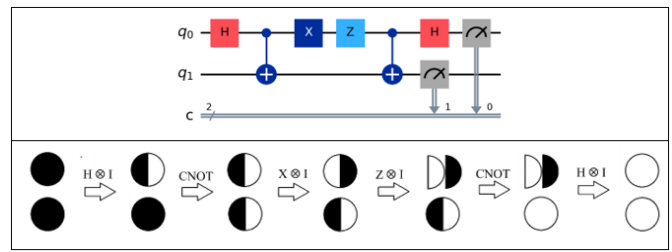

**Fig.8** Quantum circuit sending |11⟩ message by super dense coding (top) and solution of the circuit with schematic representation (bottom)

### 4.3 Quantum Teleportation

The quantum circuit used to transmit a quantum state from Alice to Bob via quantum teleportation is given in Fig. 9. In the figure, $q_0$ and $q_1$ are Alice's qubits, and $q_2$ is Bob's qubit. The quantum state that Alice wants to transmit via teleportation is represented by $q_0$. Initially, all three qubits are in the |0⟩ state. First, this quantum state that Alice wants to send is randomly put into a superposition state with the H and $R_y$ structures. First, the black solid circle comes into superposition with the left side black and the right side white, and then the Ry gate changes the ratio of the black and white halves. In the figure, the fact that these halves are of different sizes is to show that the coefficients *a* and *b* are different. Thus, a random quantum state is created for the $q_0$ qubit ($a|0\rangle+b|1\rangle$)). Since no operation is applied to the other qubits until this stage, they maintain their initial state, which is the black solid circle.

In the schematic representation, the left halves of these circles representing qubits correspond to each other, while the right halves correspond to each other. It is not possible to compare the left half of one of qubit with the right of the other qubit, or vice versa. Therefore, it would not be correct to show the left half of the shape formed when the H gate is applied to q1 as black and the right half as white. If this is the case, it would mean that when $q_0$ is black, $q_1$ will also be black, and when $q_0$ is white, $q_1$ will also be white. However, no entanglement has yet been created between $q_0$ and $q_1$. Therefore, since $q_0$ can be black or white when $q_1$ is black, and $q_0$ can be black or white when $q_1$ is white, the representation in the figure is used. The upper left half of $q_1$ is black, the lower left half is white. The upper right half of $q_1$ is white, and the lower right half is black (The total ratio of white and black is still half). After this stage, the CNOT gate is applied to create entanglement between $q_1$ and $q_2$. Therefore, $q_2$ takes the same circular appearance as $q_1$entanglement between $q_1$ and $q_2$. Therefore, $q_2$ takes the same circular appearance as $q_1$.

At this stage, $q_0$ and $q_1$ are entangled by applying CNOT between them. Since the left side of $q_0$ is black, no operation is applied to the left side of $q_1$, whereas when the right side of $q_0$ is white, the right side of $q_1$ is flipped. Accordingly, the upper right half of $q_1$ is black, and the lower right half is white. Finally, in the last stage, H gate is applied to $q_0$. The upper side of the black left half becomes black, and the lower side becomes white, while the upper side of the white right side becomes black and the lower side becomes white in mirror image (-1).

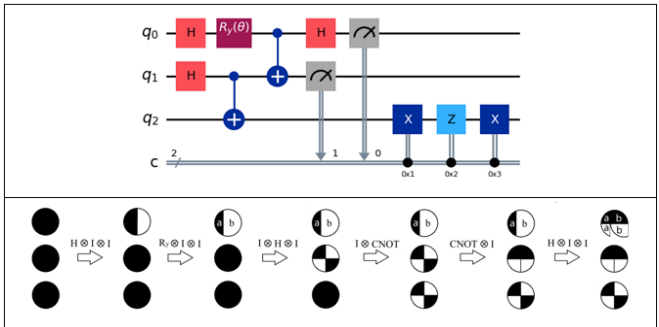

**Fig.9** Quantum teleportation circuit (top) and solution of the circuit with schematic representation (bottom)

Now Alice can measure her own qubits ($q_0$ and $q_1$). If Alice measures $q_0$ and $q_1$ as black and black respectively, as is clear from the schematic diagram, this corresponds to one of two possibilities. The first of these possibilities is the upper left of $q_0$ and the upper left of $q_1$. The second is the upper right of $q_0$ and the upper right of $q_1$. These correspond to the cases where $q_2$ is *a* times black and *b* times white, respectively. Due to the entanglement between them, the upper left corner and the upper right corners of $q_1$ and $q_2$ qubits are entangled (Fig.10.a). If Alice observes the first of her qubits, $q_0$, as black (top left) and $q_1$ as white (bottom left), then Bob's qubit state will be *a* times white (bottom left). In the other case, where $q_0$ is black (top right) and $q_1$ is white (bottom left), then Bob's qubit state will be *b* times black (bottom left) (Fig.10.b). There are two possibilities where Alice measures her qubits as white and black, respectively. The first is the case where $q_0$ is white (bottom left) and $q_1$ is black (top left). Under these conditions, Bob's qubit state will be *a* times black. Since $q_1$ and $q_2$ are entangled, we have considered the upper right side at q1 if and only if we have considered the upper right side at $q_2$. The second is the case where $q_0$ is white (bottom right) and q1 is black (top right). Under these conditions, Bob's qubit state will be -*b* times white (Fig.10.c). If Alice measures q0 and q1 as white and white respectively, Bob must obtain either *a* times white (both lower left) or -*b* times black (both lower right) (Fig.10.d).

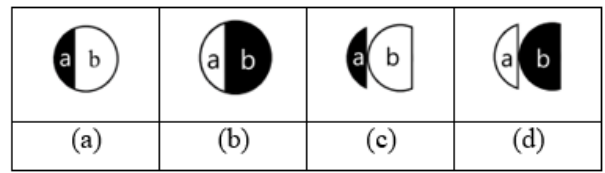

**Fig.10** Bob's qubit states corresponding to the four possibilities (a for $|00\rangle$, b for $|01\rangle$, c for $|10\rangle$, and d for $|11\rangle$) that Alice could measure her qubits with.

If Alice measures her qubits as black-black $|00\rangle$, Bob applies an I gate to his qubit and obtains the quantum teleported state (Figure 10.a). However, if Alice measures her qubits as black-white $|01\rangle$ and transmits this information to Bob, Bob applies an X gate to his qubit. In the schematic representation of Bob's qubit in Figure 10.b, the white left half of the circle will be turned black, and the black right half will be turned white. Bob will again have the correct state that Alice wants to transmit. If Alice give the information that her qubits are white-black $|10\rangle$, Bob applies a Z gate to his qubit (Figure 10.c). Thus, the black left half of the circle representing the qubit remains the same, while the white right half is mirrored, yielding the original transmitted state. If Alice measures her qubits as white-white $|11\rangle$ and passes this information to Bob, Bob first applies an X gate and then a Z gate to his own qubit. When the X gate is applied, the left half of the circle representing the qubit in Figure 10.d turns from white to black, and the right half from black to white. The Z gate applied next causes the right half to be horizontally mirrored. Bob is back to the original state in Figure 10.a.

## Conclusions

In this work, a new schematic representation method is developed to facilitate the understanding of quantum circuits and gate effects. Although the power of traditional matrix algebra and Dirac notation is indisputable, intuitive difficulties may arise in conceptual understanding and rapid problem solving. To overcome these difficulties, an innovative approach is developed that visualizes quantum states using black and white circles and represents quantum gates by visual transformations on these circles. The black circle represents the $|0\rangle$ state, the white circle represents the $|1\rangle$ state, and the half-black-half-white circle represents the superposition state. When quantum gates such as the Hadamard gate are applied, these circles are visually transformed to intuitively reflect the effect of the operation.

This schematic representation method simplifies quantum algorithms, entanglement, and multi-qubit operations transparent and simple, instead of complex mathematical operations, for understanding the basic principles of quantum mechanics. It is demonstrated in solving some basic quantum computational concepts such as Bell state generation, superdense coding, and quantum teleportation. These examples show that the developed schematic representation is a truly appropriate tool for following the stepwise evolution of quantum circuits and visually understanding quantum computational concepts.

The schematic representation method presented in this work may have important applications in quantum computing education, quantum circuit design, and rapid debugging of quantum programming frameworks. Future research may aim to apply this schematic method to different quantum algorithms and more complex quantum circuits, and integrate it into quantum computing tools. Thus, it may contribute to the development of the quantum computing field and its reach to a wider audience.